\documentclass[preprint2]{aastex}

\shorttitle{A tomographic study of the classical nova RR Pic}
\shortauthors{Ribeiro, F.M.A, Diaz, M.P.}

\begin{document}


\title{A tomographic study of the classical nova RR Pic}


\author{Fab\'{\i}ola M. A. Ribeiro and Marcos P. Diaz}
\affil{Instituto de Astronomia, Geof\'{\i}sica e Ci\^encias Atmosf\'ericas, Universidade de S\~ao Paulo,\\ r. do Mat\~ao 1226, cid. Universit\'aria, cep 05508-900, S\~ao Paulo, SP, Brazil}
\email{fabiola@astro.iag.usp.br, marcos@astro.iag.usp.br}


\begin{abstract}
We present the results of spectrophotometric observations of the old nova \\RR Pic performed in two spectral ranges, one centered in the H$\alpha$ line and other covering H$\beta$ and H$\gamma$ spectral lines. From the H$\beta$ radial velocity study we found a primary radial semi-amplitude of $K_1$ = 37(1) km s$^{-1}$ and a systemic velocity of $\gamma$ = 1.8(2) km s$^{-1}$. With this new values a mass diagram is constructed, constraining the possible mass intervals for the system. The possible orbital inclination range was restricted using the  fact that RR Pic presents shallow eclipses. A secondary mass range below the limit of a main sequence star filling its Roche lobe indicate an evolved companion. We also calculated the H$\alpha$, H$\beta$, H$\gamma$, HeI 6678 and HeII 4686 Doppler tomograms. The most conspicuous differences are found between the HeI and HeII tomograms, the former has a ring shape, while the second is filled at low velocities, suggesting that the low velocity emission is not coming from the accretion disk. Radial emissivity profiles for these lines were also derived.
\end{abstract}





\section{Introduction}

Cataclysmic Variables are close binaries systems composed by a white dwarf that accretes matter from a red dwarf or subgiant star via an accretion disk, if the magnetic field of the primary is negligible. Classical novae are eruptive cataclysmic variables with only one high amplitude outburst observed. The spectra of recent novae could show, depending of their evolutionary epoch after the eruption, a complex superposition of the spectra of the accretion disk, shell permitted and forbidden lines and eventually signatures of the secondary star (see \citet{Jon31} for spectral evolution of RR Pic before the eruption). The irradiation of the secondary by the shell ionizing source may induce additional line emission \citep*{Per02} though the shielding of the white dwarf by the accretion disk should be relevant once the accretion has been reestablished.

Novae and some nova-like systems usually present intense HeII emission lines. The ratio HeII/H$\beta$ is often much smaller for quiescent dwarf novae than it is in nova remnants and nova-likes. The UX UMa type nova-likes present a variable intensity of HeII 4686, as in IX Vel where the HeII and CIII/NIII lines are present in some nights and absent in others \citep{Hes90}. It has been suggested in the past that the HeII line is not produced by viscous heating in the accretion disk, but it is a recombination line produced by photoionization in a region illuminated by the boundary-layer ionizing photons \citep{Wil80}.

RR Pic is a cataclysmic variable classified as a classical nova, with eruption recorded in 1925. \citet*{van66} observed periodic variations its light curve, and also observed eclipses which did not occur in all conjunction phases. \citet{Vog75} determined the RR Pic's orbital period from photometric observations as being 0.1450255(2) d. \citet{War86} presented RR Pic light curves from 1972 to 1984, showing an intense brightness modulation and active flickering. The presence of shallow and irregular eclipses is mentioned by \citet{War87}.

Spectroscopy of the nova shell filaments was performed by \citet*{Wil79} as a part of their study of the physical conditions in novae ejecta. They also estimated the separation between the two main knots in the shell of RR Pic, first observed by van den Boos and Finsen (cited in \citet{Jon31}). The observation of the shell knots at different epochs reveal their average expansion velocity. The last imaging of RR Pic showing the shell features and dimensions was taken by \citet*{Gil98}. They obtained a 30'' separation between two knots at opposite sides of the shell, so, we could accept this value as an approximate value of the current shell apparent dimension.

Fast photometric variability was first detected by \citet{War76} with a period of about 30 s, the variability was confirmed \citep{War81} with periodicities at 20 s and 40 s, with a more persistent period of 32 s. \citet*{Fri98} performed a wavelet transform study of the flickering of some cataclysmic variables, including RR Pic. Such a study shows that RR Pic has an intense fast photometric activity when compared to other novae. This nova is also suspected of being an intermediate polar \citep{Kub84}, on the basis of the detection of a coherent brightness modulation of about 15 s in U, B and V bands. However, \citet*{Hae85} could not confirm the existence of this period using a large photometric database.

The RR Pic's orbital period is 0.14502545(7) d, calculated by \citet{Kub84}. \citet*{Hae91} performed time resolved spectroscopy of the H$\alpha$ line, presenting the first measurements of the line profile variations with the orbital phase. \citet{Sch03} also studied the H$\alpha$ and HeI line profiles.

In this work we propose to locate and quantify the HeI, HeII and Balmer line sources as well as constrain the stellar masses in this system. The observations and data reduction are detailed in the section 2. The radial velocity study, mass constraints and Doppler tomography are shown in section 3.  A discussion of the results is made in section 4. Finally, our conclusions are outlined in section 5.

\section{Observations}


The RR Pic spectrophotometric observations were made from 2001 to 2003. The observations were performed with the 1.60 m Perkin-Elmer telescope at LNA - Brazil and with the 1.52 m ESO telescope at La Silla - Chile. In both cases we used Cassegrain spectrographs with a spectral resolution of about 2 \AA. The first observations were aimed at the H$\alpha$ line. The last ones were intended to cover H$\beta$, He II and H$\gamma$ spectral lines. For more details see journal of observations below.

\placetable{tbl-1}

The slit position angle was chosen to include a comparison star in the slit and also to be at an angle that avoids the shell bright knots (see \citet*{Gil98} for a shell image). The slit width was set to include a fraction of about 2/3 of the star seeing disk, so a negligible part of the shell emission is included in our spectra. In general, the contribution of the shell emission over the stellar profile is well subtracted by interpolating the local background along the spatial direction.

The observations were bracketed by arc lamp exposures in a regular basis to allow good wavelength calibration of the spectra. The interval between consecutive lamp exposures was estimated considering the spectrograph mechanical flexure maps, aiming to minimize its effects in the derived velocities. The dispersion solutions were interpolated in airmass for each target exposure.

Differential spectrophotometry was performed, using the integrated flux from the slit comparison star. In order to perform the absolute flux calibration of all the spectra, tertiary spectrophotometric standard stars were also observed \citep{Ham94}. Wide slit observations of the slit comparison star were made to correct our spectra from slitlosses and from differential atmospheric dispersion effects.

All the data reduction was made using standard IRAF\footnote[1]{IRAF is distributed by the National Optical Astronomy Observatories, which are operated by the Association of Universities for Research in Astronomy, Inc., under cooperative agreement with the National Science Foundation.} reduction procedures. The images were bias subtracted and corrected from flatfield. Then the spectra were extracted using optimal extraction algorithm \citep{Hor86}, calibrated in wavelength and flux. The spectra in the red range were also corrected from telluric absorption effects in the vicinity of H$\alpha$ line using a scaled telluric absorption template with the same spectral resolution.


 

\section{Results}

\subsection{Spectral Features}

We present in figure 1\notetoeditor{figure 1 must be placed as a double column figure} the average spectra in the red region, covering H$\alpha$ and HeI 6678 lines and in the blue region, covering H$\beta$, H$\gamma$ and HeII 4686 lines. The Balmer and HeI lines seem double peaked while the HeII appears single peaked with extended wings. Only the most intense lines can be used for Doppler tomography. We can see that near HeII there is a blend of CIII and NIII lines. Unfortunely, the blending between them is too severe and prevent the Doppler mapping of these lines. No absorption lines from the secondary could be detected.

\begin{figure}
\epsscale{2}
\plotone{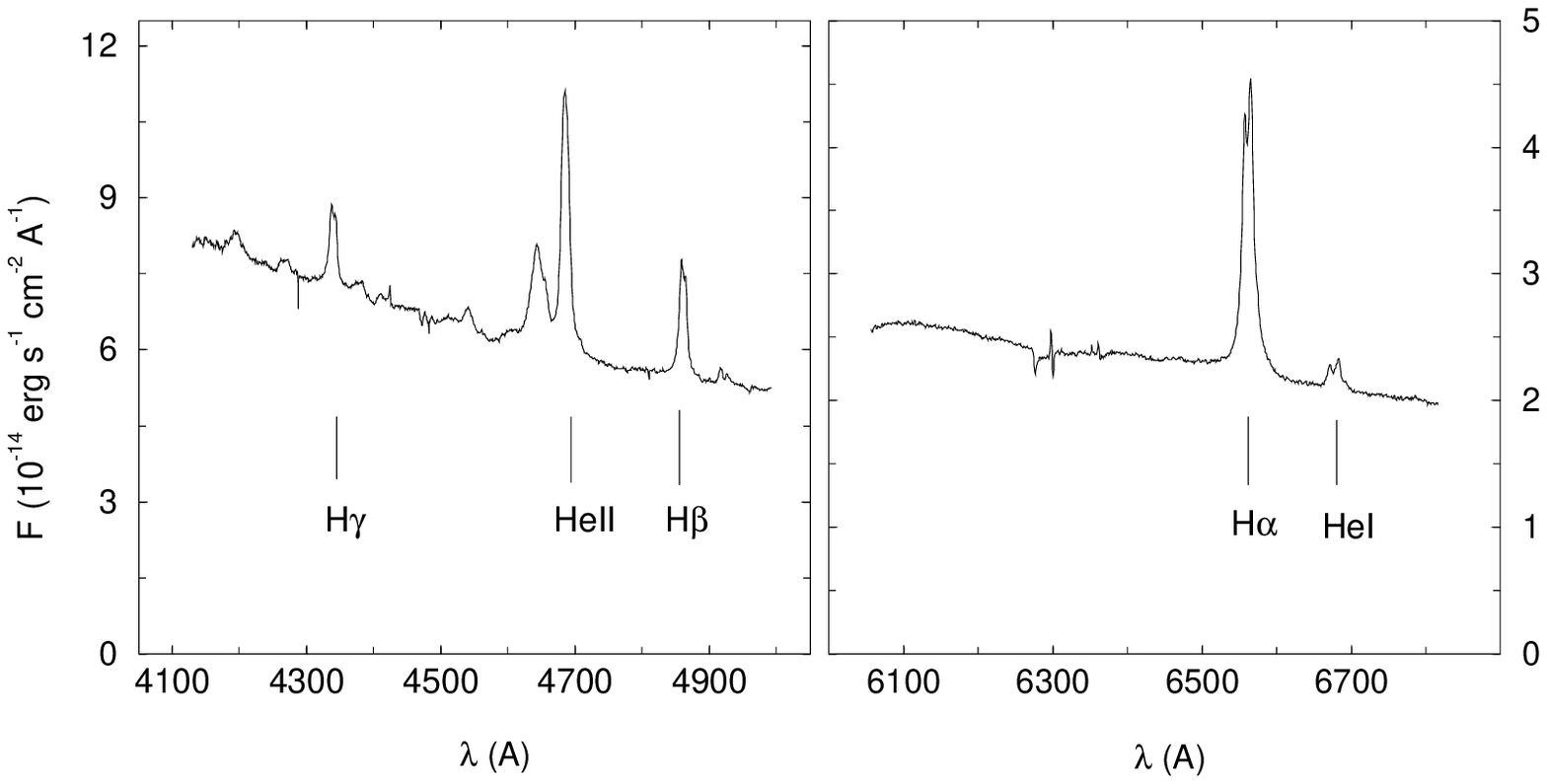}
\figcaption{Average spectra, in red region (right) and blue region (left). The lines used for Doppler tomography are identified. The features near H$\gamma$ line are the result of a bad column in the CCD used at ESO telescope. The feature at 6277 \AA~is a telluric O$_2$ absorption band. The feature at 6300 \AA~is a telluric [OI] emission due to bad sky subtraction.\label{fig1}}
\end{figure}

\placefigure{fig1}

\subsection{Line Profile and Radial Velocity Study}

We can see in figure 1 that the HeII line is blended with the CIII/NIII complex, but this blending does not compromise too much the HeII blue wing profile, so we can simply limit the maximum velocity used in HeII radial velocity study to avoid the nearby lines. The maximum velocity was fixed at 1000 km s$^{-1}$.

The spectra were binned in phase boxes, using the orbital period proposed by \citet{Kub84} and our spectroscopic conjunction phase (see fig. 4), the spectra were binned in phase boxes. From the phase diagrams it could be noticed that the lines do not present a large oscillation around the rest wavelength, indicating a low primary's radial velocity semi-amplitude. As we go to the line wings we can see that the oscillation around the rest wavelength becomes more noticeable. No emission is found above 1200 km s$^{-1}$.

The H$\alpha$ line profile does not show a clear single peaked profile at any orbital phase (fig. 2)\notetoeditor{figure 2 must be placed as a single column figure}. The H$\alpha$ line presents a more intense emission near $\phi$ = 0. The HeI line is also more intense near phase $\phi$ = 0, has a more structured shape compared to the H$\alpha$ line, and also shows emission at larger velocities. For H$\beta$, H$\gamma$ and HeII, an increase of the line intensity is found near phase 0.6. The phase sampling of our data was verified, confirming that this intensity increase could not be caused by an irregular phase coverage.

\begin{figure}
\epsscale{.9}
\plotone{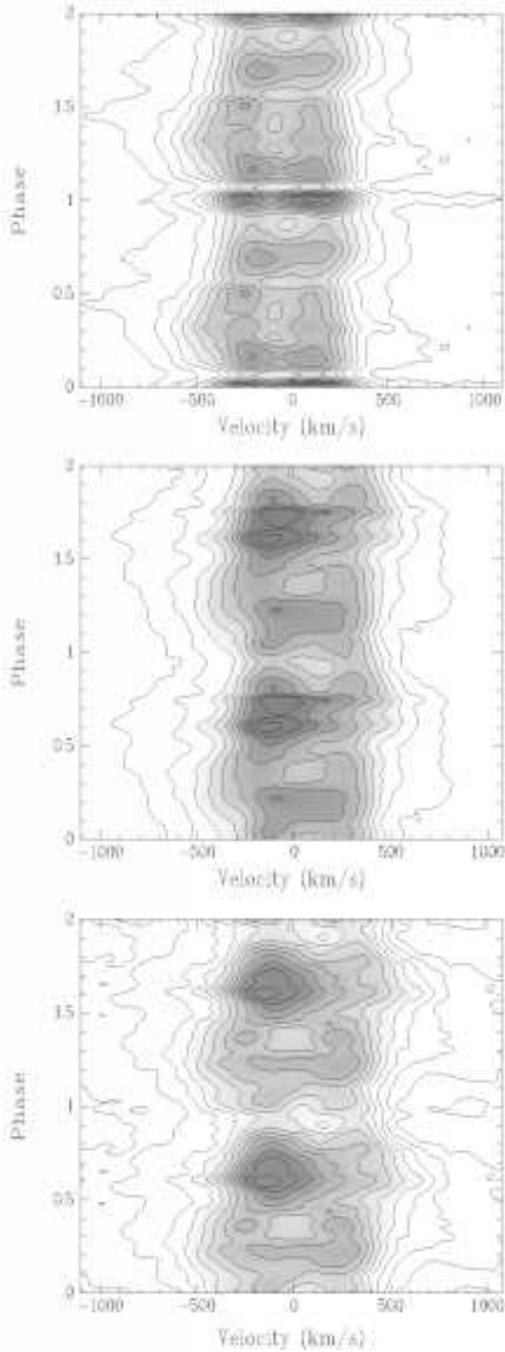}
\figcaption{Phase diagrams of the H$\alpha$ (upper panel), H$\beta$ (middle panel) and H$\gamma$ (lower panel) continuum subtracted lines. The grayscale was set to properly show the main features in each graph. Contour lines are regularly spaced from zero to the maximum intensity.\label{fig2}}
\end{figure}

\placefigure{fig2}

\begin{figure}
\epsscale{1}
\plotone{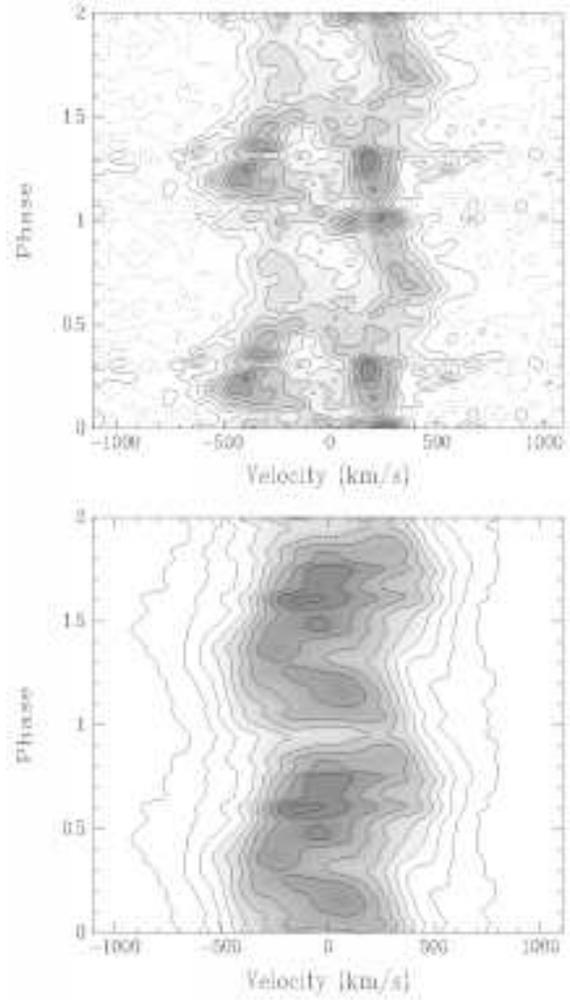}
\figcaption{Same as fig. 2 for HeI 6678 (left) and HeII 4686 (right).\label{fig3}}
\end{figure}

\placefigure{fig3}

The HeI and HeII (fig. 3)\notetoeditor{figure 3 must be placed as a single column figure} phase diagrams have different shapes, but its important to recall that they were observed at different epochs. The H$\alpha$ phase diagram is slightly different from those of H$\beta$ and H$\gamma$, and they were also derived from data taken at different dates.

The H$\beta$ line data were used to estimate the primary's radial velocity semi-amplitude $K_1$. This line was chosen because it is one of the most intense in our spectra, it is not blended and also because there are more independent observations in the blue dataset than in the red set. The diagnostic diagram (fig. 4)\notetoeditor{figure 4 must be placed as a single column figure} is built using radial velocities derived by convolving the line profiles with a double Gaussian mask (\citet*{Sch80}, hereafter SY). A mask with 50 km s$^{-1}$ (FWHM) Gaussians was applied. Different values of the Gaussian's half-separation were used in order to sample different projected velocities in the line profile. The Gaussian half-separation velocity ``$|V|$'' appear in the diagram as the horizontal axis. For each half-separation a radial velocity curve is obtained, fitted with a periodic function and the parameters of such this fit are given in the y axis of the panels in the diagnostic diagram. The phase scale is given by the spectroscopic inferior conjunction of the secondary, i.e. the timing for positive to negative crossing of the line wing radial velocity curves. The ``best'' value of $K_1$ found in the line wings is the one found at a minimum RMS, on a plateau in the $K_1$ curve . In order to estimate $K_1$, an average of the $K_1$ values for velocities ranging from 466 km s$^{-1}$ to 605 km s$^{-1}$ was made, obtaining $K_1$ = 37(1) km s$^{-1}$. A systemic velocity $\gamma$ = 1.8(2) km s$^{-1}$ and the spectroscopic secondary conjunction phase $\phi_0$ = 2452295.7744(3) HJD were estimated using the same range in ``$|V|$''. One can see that the diagnostic diagram is well behaved, for such a large velocity range in the line wings we have a small value of the systemic velocity  and also a small zero phase variation.

The problem of estimating the white dwarf orbital velocity using emission lines in CVs is a classical issue. It is, of course, desirable that the measurement of $K_1$ is performed at the highest possible velocity in the wings. We simply expect that the high velocity gradient in the disk suffer from less anisotropies. A plateau in the diagnostic curve is also expected if there are no anisotropies. Choosing $K_1$ from a steep diagnostic curve is much more complicated. Possibly, different authors with different S/N ratios in the line wings will find different $K_1$ values. For instance, if it is a raising $K_1$ versus $|V|$ diagnostic curve, the observer who has the best S/N will derive the largest $K_1$ values.

The uncertainty in $K_1$ is derived from the dispersion between $K_1$ values in the diagnostic diagram plateau. The sinusoid amplitude fitting uncertainty of individual $K_1$ points is of the same order (2 km s$^{-1}$). However, both are just formal errors. The uncertainty in $K_1$ propagates to the derived mass ranges, so these ranges are just formal as well. They do not include the effect that measured $K_1$ velocities in CVs are possibly not the white dwarf orbital velocities. We have cross checked our H$\beta$ wing velocities with H$\alpha$ measurements using the same convolution masks and comparable results were found. Again, H$\beta$ and H$\alpha$ were measured at different epochs.

In the next sections the $K_1$, $\gamma$ and phase reference values found here will be adopted.

\begin{figure}
\epsscale{.8}
\plotone{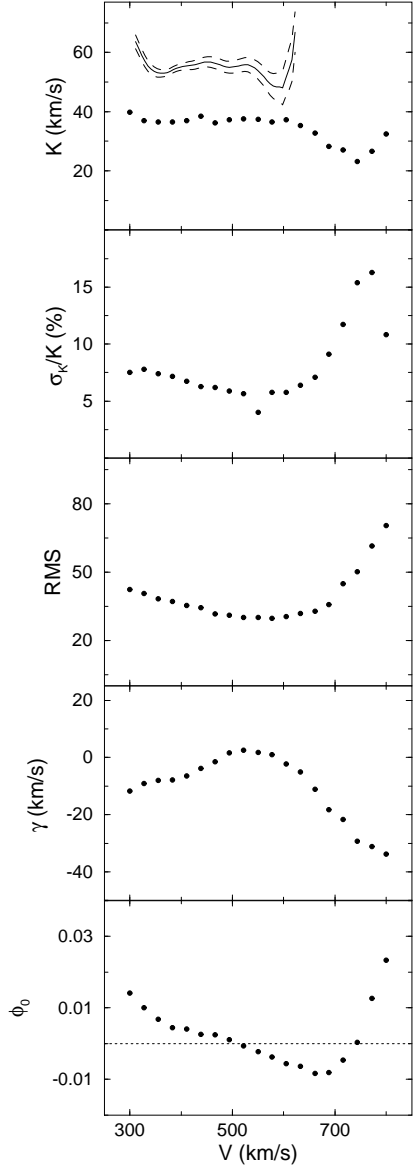}
\figcaption{Diagnostic diagram for the H$\beta$ line. The x-axis is the half-separation of the Gaussians in the SY method. The panels from top to bottom show the H$\beta$ radial velocity semi-amplitude, its relative error, the RMS of the radial velocity curve fit, the systemic velocity and the phase shift. The solid and dashed curves in the first panel represents the radial velocity semi-amplitude obtained from the H$\beta$ Doppler tomogram and its 1$\sigma$ uncertainty.\label{fig4}}
\end{figure}

\placefigure{fig4}

Another way of obtaining the primary's radial velocity semi-amplitude is from a Doppler tomogram centered at the system's center of mass. Circular isophotes with increasing radius (or velocity modulus) were fitted to the H$\beta$ tomogram constructed as described above. The inner isophotes will follow the bright features in the Doppler map but the outer will tend to trace the intrinsic high velocity emission in the disk. In the first panel on figure 4 the continuous line represents the $K_1$ values obtained from the displacement of each isophote center y-axis origin. It is found that the value of $K_1$ obtained from this method is larger than that obtained from double Gaussian mask convolution for $|V| <$ 700 km s$^{-1}$. They become marginally compatible with the SY plateau only at $|V| >$ 700 km s$^{-1}$. This difference could be explained if we consider that in the SY method we have sampled the emission from regions with different intrinsic velocities that yields the same projected velocity inside the Gaussian mask. In the next steps we will use the value of $K_1$ obtained from the double Gaussian method. The SY method was preferred to the tomogram isophote fitting in the particular case of RR Pic because $K_1$ is much smaller than the tomogram FWHM resolution. The presence of asymmetries in the brightness distribution at high velocities may impact the determination of isophote centers and, consequently, produce a relatively large systematic effect on $K_1$.

A radial velocity study of RR Pic was also performed by \citet{Sch03}, using H$\alpha$ and HeI 6678 spectral lines. These authors estimated a primary radial semi-amplitude of $K_1 \sim$ 170 km s$^{-1}$. This value could not be confirmed by our measurements of the H$\beta$ line wings. \citet{Sch03} have measured $K_1$ at wing velocities of $\sim$ 800 km s$^{-1}$. There is almost no signal at such velocities. This can be verified by inspecting their tomograms and phase maps. In addition, both their radial velocity analysis and Doppler tomography were computed using a rather small number of independent phases (19 spectra).

\subsection{Constraints on Stellar Masses}

Considering the geometry of the system, the presence of shallow eclipses, a maximum disk radius \citep{Pac77} and the volume radius of the secondary \citep{Egg83} we find a possible inclination range $60 \degr < i < 80 \degr$. This result is similar to that obtained by \citet{Hor85} considering the eclipse of the central region of the disk.

\begin{figure}
\epsscale{1.8}
\plotone{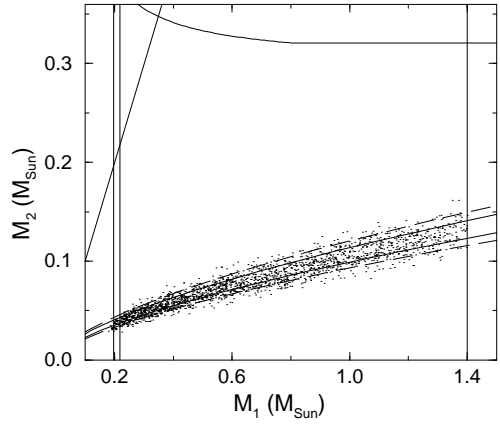}
\figcaption{RR Pic mass diagram. The vertical lines are lower and upper limits for the white dwarf mass (see text), the diagonal line at the left is the limit for stable mass accretion and the curve at the top is the upper mass limit for a main sequence secondary filling its Roche lobe (\citet{Pat84}, eq. 7). The solid lines at the lower part of the graph limit the stellar masses, considering inclinations between 60$\degr$ and 80$\degr$ and $K_1$ = 37 km s$^{-1}$, the dashed lines represent this region considering 1$\sigma$ uncertainty. The dots were obtained from a bootstrapping simulation (see text).\label{fig5}}
\end{figure}

\placefigure{fig5}

In figure 5\notetoeditor{figure 5 must be placed as a single column figure} we present the mass diagram for RR Pic. A lower limit of 0.2 M$_\sun$ for the primary mass is derived from the H$\beta$ line FWZI. The upper limit for its mass was fixed at 1.4 M$_\sun$. In addition, the secondary mass must be roughly equal or lower than the primary's for a stable accretion regime. The secondary also needs to have a mass below the limit for a main sequence secondary filling its Roche lobe \citep{Pat84}.

Using broad limits on M$_1$ ($0.2 < M_1 < 1.4~M_\sun$), in addition to the previously derived value K$_1$ = 37 km s$^{-1}$ and inclination range ($60\degr < i < 80\degr$), we found from the mass function that $0.09 < q < 0.2$. A secondary's radial velocity semi-amplitude $K_2$ ranging from 200 km s$^{-1}$ to  400 km s$^{-1}$ is also derived.

A bootstrapping simulation \citep{Horb86} was performed aiming to confirm the mass intervals given by the mass diagram. An inclination range of 60$\degr$ to 80$\degr$ with a flat error distribution and a $K_1$ value of 37(1) km s$^{-1}$ were used in this simulation. The results of this simulation are shown in figure 5. One can see from this same figure that the secondary must have a mass smaller than  0.15 M$_\sun$. This indicates a low mass ratio ($q = M_2 / M_1$) for the system. Calculating the secondary's mean density from the equation 1 \citep{War95}, we find $\bar\rho$ = 8.8 g cm$^{-3}$. From this result one finds that the spectral type of the secondary should be near M5 if it is a main sequence star \citep{All00}.

\begin{equation}
\bar \rho_2=107 P_{orb}^{-2}(h)~~ g~cm^{-3}
\end{equation}

The fact that the secondary has a mass much smaller than a main sequence star  suggests that the secondary may have evolved from the main sequence. One can also constrain the white dwarf mass considering the fact that RR Pic had a moderately fast nova outburst. The white dwarf mass is probably greater than 0.6 M$_\sun$, as expected from classical novae outburst models \citep{Sta89}. In addition, the white dwarf mass is probably not too close to the Chandrasekhar limit, since no recurrent nova outbursts were observed in RR Pic over the last 80 years \citep{Web87}.
In the next steps we will use the masses given by the center of the most probable mass region in the mass diagram: $M_1$ = 1 M$_\sun$ and $M_2$ = 0.1 M$_\sun$.

\subsection{Doppler Tomography}

The Doppler tomography method was applied to the brightest emission lines, obtaining Doppler maps with origin at the system's center of mass. Before interpreting the Doppler tomograms is important to notice that the H$\alpha$ HeI observations and the H$\beta$ H$\gamma$ HeII data were taken at different epochs. This fact implies that tomograms from each dataset should be compared only with tomograms from the same dataset. The position of the secondary and the binary center of mass, as well the primary and $L_1$ Lagrange's point were plotted for a 90$\degr$ orbital inclination. These points will be displaced in their y-values for a different orbital inclination. The self-consistency of the Doppler maps could be verified comparing the tomogram projections with the observed spectra at its correspondent orbital phase. The projections have a good agreement with their equivalent line profiles, both in flux and in line profile shape.

The H$\alpha$ tomogram (fig. 6)\notetoeditor{figure 6 must be placed as a single column figure} presents a ring shape, as expected for an accretion disk reconstruction, where there is a low velocity limit in the outer disk radius and an observational high velocity limit at the inner disk. The H$\alpha$ tomogram presents most intense emission in the $(-V_x, +V_y)$ and $(-V_x, -V_y)$ quadrants. In contrast, the H$\beta$ and H$\gamma$ Doppler tomograms also present the signature of a disk, but with a weaker emission in the $(+V_x, -V_y)$ quadrant. When compared to the H$\alpha$ map an emission deficit in the $(+V_x, +V_y)$ quadrant is verified in the upper Balmer transitions.

\begin{figure}
\epsscale{.8}
\plotone{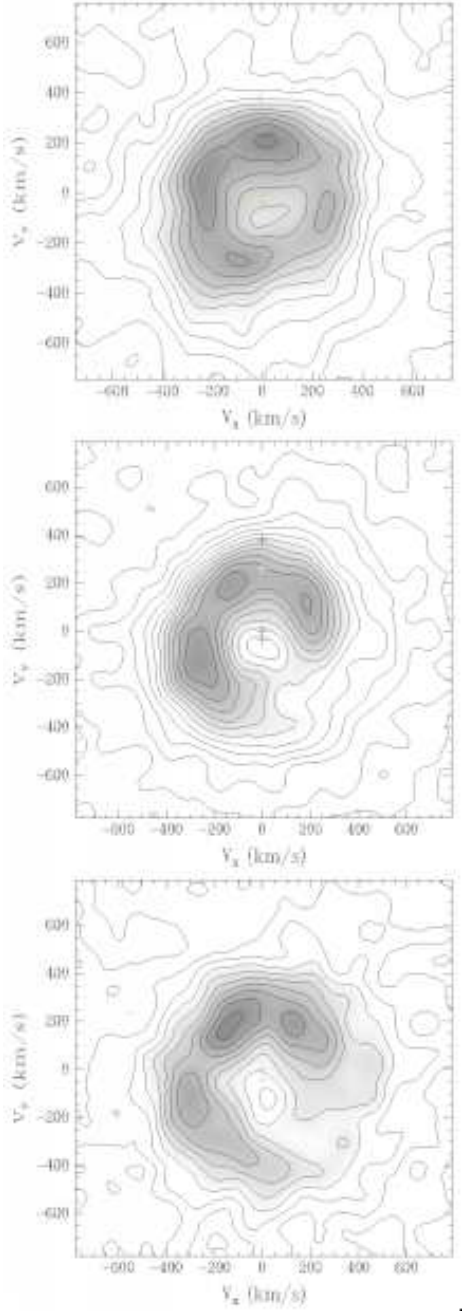}
\figcaption{H$\alpha$ (upper panel), H$\beta$ (middle panel) and H$\gamma$ (lower panel) Doppler tomograms. Observed velocities are quoted. The ``+'' signs indicate, from up to bottom, the secondary's center of mass, the $L_1$ Lagrangian point, the system's center of mass and the primary's center of mass, using M$_1$ = 1 M$_\sun$ and M$_2$ = 0.1 M$_\sun$ (as described in section 3.3).\label{fig6}}
\end{figure}

\placefigure{fig6}

\begin{figure}
\epsscale{1}
\plotone{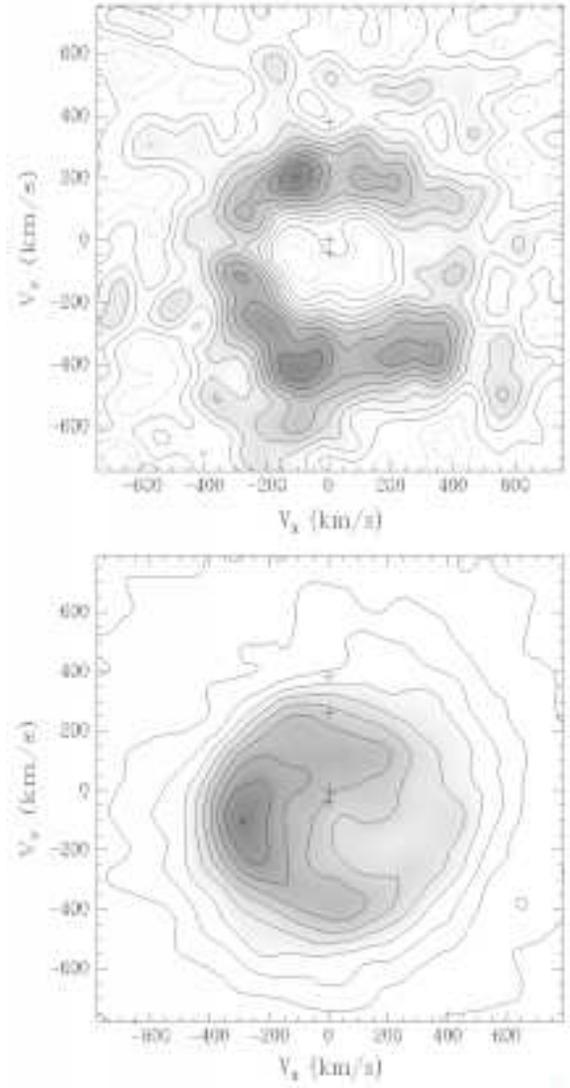}
\figcaption{Same as fig. 6 for HeI 6678 (top) and HeII 4686 (bottom).\label{fig7}}
\end{figure}

\placefigure{fig7}

The HeI Doppler tomogram (fig. 7)\notetoeditor{figure 7 must be placed as a single column figure} presents a ring shape, but it shows the inner ring radius at velocities greater than those found in the H$\alpha$ tomogram, suggesting that the HeI emission is enhanced in the inner disk region. An emission enhancement in the lower part of the tomogram is also seen. The HeI Doppler tomogram is noisier than the other tomograms because the HeI 6678 line is much fainter than the other ones.

The HeII 4686 Doppler tomogram (fig. 7) presents a distinct behavior when compared to Balmer and HeI tomograms, showing emission at very low velocities. This low velocity emission can be explained by the presence of a wind coming from the accretion disk, or by the emission from stationary material inside the Roche lobe. This emission could also be associated with gas spilling over the disk. The line production mechanism for such vertically extended gas distribution may be recombination as the gas is easily irradiated by the inner disk region and boundary layer. Following this interpretation, the region of enhanced emission in the $-V_x$ region of the HeII tomogram could be associated with the hot spot, where the stream hits the accretion disk. An enhanced  emission in this same quadrant could be also seen in H$\beta$ tomogram.

\subsection{Accretion Disk's Radial Emissivity Profiles} 

The disk radial emissivity profiles obtained from H$\alpha$, H$\beta$, H$\gamma$, HeI 6678 and HeII 4686 spectral lines are presented in figure 8\notetoeditor{figure 8 must be placed as a double column figure}. The Doppler maps previously discussed are centered in the system's center of mass. By including the primary's radial velocity, one may shift the Doppler tomograms to be centered at the white dwarf. The radial disk emissivity profile is estimated from these tomograms by calculating the mode of the intensity over concentric rings centered at the origin. The mode was chosen as the statistical estimator, allowing us to obtain the emissivity of the disk disregarding large emission anisotropies. Using a primary's mass of 1 M$_\sun$, the Doppler tomograms were converted from velocity space to position space using a Keplerian velocity law. The radial emissivity profiles are corrected for reddening considering the color excess $E(B-V)=0.02$ \citep*{Bru94} and R = 3.1. The radial emissivity profile inclinations obtained are -1.5 for H$\alpha$ and H$\beta$, -1.7 for H$\gamma$ and HeII and -1.9 for HeI, the error of these values is about 0.1.

\begin{figure}
\epsscale{2}
\plotone{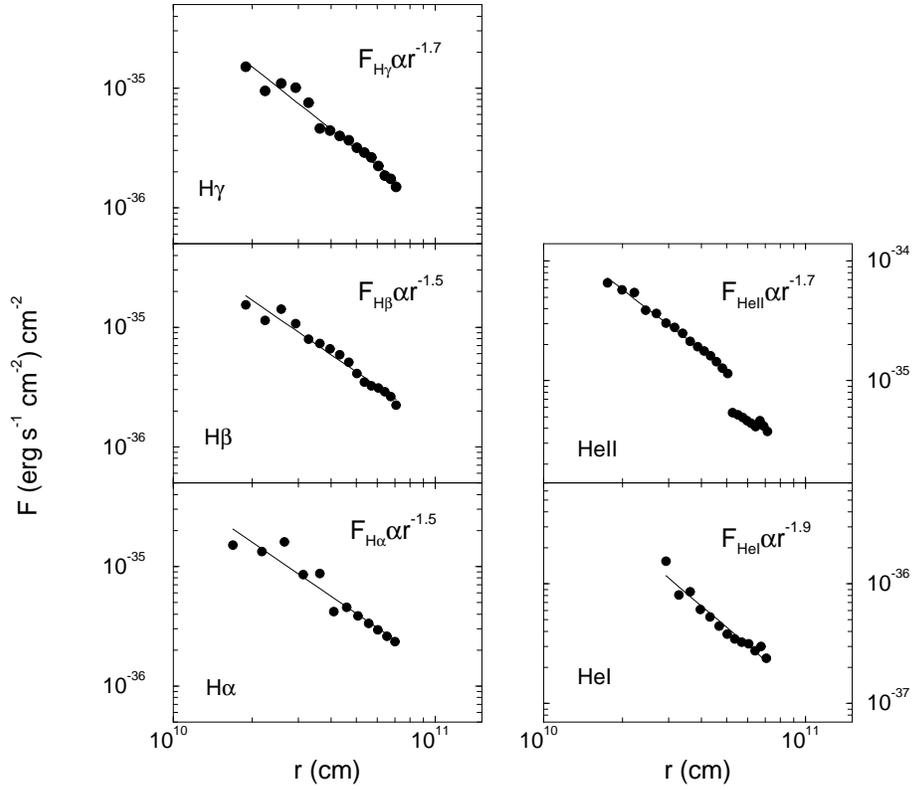}
\figcaption{Disk radial emissivity profiles for the H$\alpha$, H$\beta$, H$\gamma$, HeI 6678 and HeII 4686 spectral lines assuming 1.0 M$_\sun$ as the primary's mass. The flux scale of the graphs are different, but they have all a 2 dex range. The power law index for HeII was derived from the inner points. The units are the observed flux corrected for reddening per unit area of the disk.\label{fig8}}
\end{figure}

\placefigure{fig8}

The HeII radial emissivity profile shows two subsets of points with different behavior and inclinations. The discontinuity between these two subsets is at about 500 km s$^{-1}$, so this difference could not be attributed to a blending effect with the CIII/NIII complex, since blending effects should only appear at velocities greater than 1000 km s$^{-1}$.

From figure 8 it can be seen that the emission inclinations for H$\alpha$ and H$\beta$ lines are similar, and that the profile seems steeper for the HeI and HeII emission. Note however, that if there is a wind contribution in HeII it will also be present in the radial emissivity profile, so this disk emission profile may be contaminated by a wind component. To convert the Doppler tomograms to position space we assumed a Keplerian velocity law, which is not necessarily true, introducing additional errors in the radial emissivity curves. The emission power law index depends on the mass of the primary in the sense that the power law becomes steeper with increasing mass. Hence, no absolute value of the power law index can be given. However, as the change of the power law index with the primary's mass must be the same for all lines, the fact that one line emission is more centrally concentrated than other line emission is independent of the primary's mass.

\section{Discussion}

Until today, there is no well established photometric ephemeris for RR Pic in the literature. While the presence of a grazing eclipse seems to be confirmed, there is no published O-C diagram for these eclipses. As far as we understand, the phasing adopted by \citet{Sch05} is based on a single feature in \citet{War86} light curves, which was interpreted as the grazing eclipse. If correct, this ephemeris would result in a spectroscopic phase offset of 0.17. Since RR Pic presents high amplitude flickering, no fiducial phasing from grazing eclipses can be firmly established without combining several eclipse light curves and analyzing their phase residuals. Therefore, a discussion about the presence of a spectroscopic phase shift (as observed in several CVs) awaits a better definition of the eclipse ephemeris.

The secondary's mass obtained from the mass diagram is significantly smaller than the limit given by a main sequence star filling its Roche lobe. One interpretation that directly arises from this result is that the secondary could be more evolved than a main sequence star. No absorption features from the secondary that could support such hypothesis are found in our spectra. \citet*{Har05} performed K-band infrared spectroscopy of RR Pic and they have not detected any obvious absorption feature from the secondary either, so the secondary's spectral type remains elusive.

From the radial emission profiles in fig. 8 it is possible to verify that the HeI  emission is more centrally concentrated than the Balmer and HeII emission. We can draw a preliminary discussion comparing the RR Pic radial emissivity profiles with others estimatives in the literature. The RR Pic radial emission profiles inclinations are smaller than those of other systems. \citet*{Dia03} found an radial emissivity profile inclination of -2.1 for H$\alpha$ line and -2.4 for the HeI 6678 line for V841 Oph. \citet*{Dia99} found -2.3 for H$\beta$ line and -2.9 for HeII for V347 Pup. In this study we obtained -1.5 for H$\alpha$ and H$\beta$ lines, -1.7 for H$\gamma$ and HeII 4686 lines and -1.9 for HeI 6678 line, the errors are about 0.1. From these values we conclude that the line emission is less concentrated in RR Pic (assuming a 1 M$_\sun$ white dwarf) than it is in V841 Oph and V347 Pup. As the power law index increases as the white dwarf mass decreases, in order to reach the H$\beta$ power law index found in V347 Pup, RR Pic must have a white dwarf with approximately 0.3 solar masses, which seems extremely unlikely. The discontinuity found in the HeII emission profile may be regarded as an evidence of non-Keplerian motion or wind emission in the HeII line source regions.

One may ask why we don't observe the diffuse emission in the Balmer tomograms as observed in the HeII. The diffuse component is also present in Balmer lines, but for this lines the disk emission may be more intense than the diffuse component. In the HeII case, the diffuse emission, produced by recombination, may be dominant when compared to the emission originated in the accretion disk. 

\section{Conclusions}

A radial velocity study of the RR Pic system was performed using extensive spectrophotometric observations. A value for the primary's radial velocity semi-amplitude of 37(1) km s$^{-1}$, considerably smaller than the value of about 170 km s$^{-1}$ given by \citet{Sch03} was found. This small primary's radial velocity implies a secondary star mass below 0.16 M$_\sun$. This mass estimative is approximately half of the limiting mass of a main sequence star filling its Roche lobe, which points out to an evolved companion star. The primary's mass could not be constrained due to the absence of secondary's photospheric lines. From the fact that RR Pic presents shallow eclipses, the system's orbital inclination was constrained to a interval between 60$\degr$ and 80$\degr$. The mass ratio $q$ could be constrained in the wide interval between 0.09 and 0.2.

The H$\alpha$ and HeI Doppler images of RR Pic show a clear ring signature.
Furthermore, the H$\beta$ and H$\gamma$ Doppler maps present ring shaped structures, while the HeII map show an enhanced emission at low velocities, indicating that this high ionization line is produced in velocity field that is different from the disk. Radial emissivity profiles were obtained from the tomograms, indicating a more concentrated emission for H$\gamma$ line than for the H$\alpha$ and H$\beta$, and a more concentrated emission from HeI while compared to HeII. In addition, the RR Pic disk may present less radially concentrated emissivity profiles, when compared to other novae and nova-like. However, the emission distribution in other quiescent disks should be derived in order to explore its correlation with other
properties of the binary system.



\acknowledgments

This work is based on data obtained at LNA/CNPq and La Silla/ESO observatories. F.M.A.R is grateful from support from FAPESP fellowship 01/07078-8. MD acknowledges the support by CNPq under grant \#301029.

\clearpage





\clearpage

\clearpage

\clearpage

\begin{table}
\begin{center}
\caption{Journal of Observations\label{tbl-1}}
\begin{tabular}{rccccc}
\tableline\tableline
Observation date & Telescope & $\lambda_{c}$ (\AA) & Exposure & Number & Number\\
 & & & time (s) & of spectra & of cycles\\
\tableline
17 Jan 2001	& 1.60m LNA & 6560 & 100 & 24  & 0.3 \\
18 Jan 2001	& 1.60m LNA & 6560 & 100 & 114 & 1.6 \\
22 Mar 2001	& 1.60m LNA & 6560 & 70  & 41  & 0.3 \\
21 Jan 2002	& 1.52m ESO & 4550 & 180 & 74  & 1.6 \\
22 Jan 2002	& 1.52m ESO & 4550 & 180 & 78  & 1.5 \\
16 Mar 2002	& 1.52m ESO & 4550 & 180 & 50  & 1.3 \\
17 Mar 2002	& 1.52m ESO & 4550 & 180 & 78  & 1.6 \\
31 Mar 2003	& 1.60m LNA & 4600 & 160 & 14  & 0.2 \\
01 Apr 2003	& 1.60m LNA & 4600 & 160 & 12  & 0.6 \\
02 Apr 2003	& 1.60m LNA & 4600 & 160 & 82  & 1.4 \\
03 Apr 2003	& 1.60m LNA & 4600 & 160 & 20  & 0.9 \\
\tableline
\end{tabular}
\end{center}
\end{table}

\end{document}